\documentclass[twocolumn,showpacs,superscriptaddress,preprintnumbers]{revtex4}
\usepackage{amssymb}
\usepackage{graphicx}
\usepackage{dcolumn}
\usepackage{bm}

\begin{document}

\title{Dynamics of slow light and light storage in a Doppler-broadened
electromagnetically induced transparency medium: A numerical approach}
\date{December 7, 2010}
\author{Shih-Wei Su}
\affiliation{Department of Physics, National Tsing Hua University, Hsinchu 30013, Taiwan}

\author{Yi-Hsin Chen}
\affiliation{Department of Physics, National Tsing Hua University, Hsinchu 30013, Taiwan}
\author{Shih-Chuan Gou}
\email{scgou@cc.ncue.edu.tw}
\affiliation{Department of Physics, National Changhua University of Education, Changhua
50058, Taiwan}
\author{Tzyy-Leng Horng}
\affiliation{Department of Applied Mathematics, Feng Chia University, Taichung 40074,
Taiwan}
\author{Ite A. Yu}
\email{yu@phys.nthu.edu.tw}
\affiliation{Department of Physics, National Tsing Hua University, Hsinchu 30013, Taiwan}

\begin{abstract}
We present a numerical scheme to study the dynamics of slow light and light
storage in an eletromagnetically induced transparency (EIT) medium at finite
temperatures. Allowing for the moitonal coupling, we derive a set of coupled
Schr\"{o}dinger equations describing a boosted closed 3-level EIT system
according to the principle of Galilean relativity. The dynamics of a
uniformly moving EIT medium can thus be determined by numerically
integrating the coupled Schr\"{o}dinger equations for atoms plus one
ancillary Maxwell-Schr\"{o}dinger equation for the probe pulse. The central
idea of this work rests on the assumption that the loss of ground-state
coherence at finite temperatures can be ascribed to the incoherent
superposition of density matrices representing the EIT systems with various
velocities. Close agreements are demonstrated in comparing the numerical
results with the experimental data for both slow light and light storage. In
particular, the distinct characters featuring the decay of ground-state
coherence can be well verified for slow light and light storage. This
warrants that the current scheme can be applied to determine the decaying
profile of the ground-state coherence as well as the temperature of the EIT
medium.
\end{abstract}

\maketitle

\address{$^{1}$Department of Appled Mathematics, Feng Chia University, Taichung 40724,\\
Taiwan\\
$^{2}$Department of Physics, National Changhua University of Education,\\
Changhua 50058, Taiwan\\
$^{3}$Department of Mathematics, National Taiwan University, Taipei 106,\\
Taiwan}


\section{Introduction}

The effect of electromagnetically induced transparency (EIT) is a nonlinear
optical phenomenon which renders an opaque medium transparent at a certain
frequency by exciting it with an electromagnetic field \cite{Marangos} . This
effect can occur generally in a 3-level atomic system, where the atomic
states are coherently prepared by external laser fields. For example, in a $%
\Lambda $-type system (see Fig.1), there are two dipole-allowed transitions $%
\left\vert 1\right\rangle \leftrightarrow \left\vert 3\right\rangle $ and $%
\left\vert 2\right\rangle \leftrightarrow \left\vert 3\right\rangle $ which
are excited by a weak probing field, and a strong coupling field,
respectively. When EIT occurs, destructive interference among different
transition pathways suppresses the transition probability between $%
\left\vert 1\right\rangle \leftrightarrow \left\vert 3\right\rangle $,
leading to a spectrally sharp dip in the absorption spectrum. The
corresponding steep dispersion within the transparency window results in a
large reduction in the group velocity of light. As the steepness depends on
the intensity of coupling field, EIT thus provides an effective and
convenient mechanism for slowing down the light in a controllable fashion.
The slow light (SL) arising from the EIT effect greatly enhances the
nonlinear susceptibility and makes the low-light level nonlinear optics
possible \cite{Harris2} . The first experimental demonstration of SL produced
by EIT was made with high-power pulsed lasers interacting with a Sr vapor
\cite{Boller,Field} . In 1999, a dramatic reduction of the group velocity
down\ to 17m/s was demonstrated by Hau \textit{et al} . by using a
Bose-Einstein condensate of Na atoms \cite{Hau} . An ultraslow group velocity
of 8m/s were later observed in a buffer-gas cell of hot Rb atoms by Budke
\textit{et al} . \cite{Bukder} .

In a lossless, passive sample, the reduction of the propagation velocity
implies a temporary transfer of electromagnetic excitations into the medium.
With EIT, the light pulse can even be completely stopped and stored in the
medium by adiabatically switching off the coupling field and subsequently
retrieved from the medium by the reverse process while the probe pulse is
entirely within the sample making its way through \cite%
{Marangos,Bukder,Fleischhauer,Phillips} . Such storage and retrieval of
photonic information are essentially reversible as they are resulted by the
coherent transfer of the quantum state of light into the quantum coherence
of the two ground states $\left\vert 1\right\rangle \ $and $\left\vert
2\right\rangle $. The light storage (LS) was experimentally demonstrated by
several groups with various schemes \cite%
{Liu,Phillips,Kocharovskaya,Turukhin} , which promises to be applied in the
processing of quantum information, especially in the implementation of
quantum storage devices, logical gates and generation of photonic qubits
\cite{Duan,Chaneliere,Vewinger} .

So far, except only a few experiments were demonstrated in atomic
Bose-Einstein condensates at nearly zero temperature \cite{Hau,Liu} , most
experimental studies for EIT were carried out at finite temperatures, using
either hot atoms at room temperatures \cite{Bukder,Kash} or laser-cooled
atoms at temperatures about a few hundreds of microKelvin \cite{Marangos} .
It is well-known that at finite temperatures, the Doppler-broadened medium
inevitably imposes a serious limit since the Doppler shifts caused by the
atomic motion introduce a randomization in the effective laser detunings
over the ensemble of atoms in the sample \cite{Javan} . Experimental
evidences indicate that even in the laser-cooled atoms, the decoherence due
to the atomic motion still can not be ignored readily, especially when the
applied external fields are in a counter-propagating geometry \cite%
{Lin,Peters} . Phenomenologically, the effect of Doppler broadening can be
addressed by including a relaxation term in the motion equation of $\rho
_{12}$. Here $\rho _{12}$ denotes the off-diagonal element of the density
operator which describes the quantum coherence of the two ground states $%
\left\vert 1\right\rangle \ $and $\left\vert 2\right\rangle $. Accordingly,
a decay constant $\gamma $ is thus introduced to account for the relaxation
of ground-state coherence. It should be noted that $\gamma $ is by no means
a universal constant which depends on the temperature as well as on the
experimental parameters such as the amplitudes of the driving fields and the
particle density of the medium. Normally, $\gamma $ can only be determined
by numerically fitting upon the experimental data. Thus one has to recompute
the numerical value of $\gamma $ once the experimental parameters are
altered in the new runs of experiment. In order to determine the dynamical
properties of the system without introducing $\gamma $, it is desirable to
develop a numerical scheme which is tractable and accessible to both
theorists and experimentalists, and in this paper we address such an issue.

In general, the atomic dynamics of the EIT medium can be approached by
employing the formalism of master equation in the Linblad form,
\begin{equation}
\frac{d\hat{\rho}\left( t\right) }{dt}=\frac{1}{i\hbar }[H,\hat{\rho}\left(
t\right) ]+\mathcal{D}\left( \hat{\rho}\left( t\right) \right)
\label{Linblad}
\end{equation}%
where $\hat{\rho}\left( t\right) $ is the density matrix of the 3-level
atom, $H$ is the total Hamiltonian, and $\mathcal{D}\left( \hat{\rho}\left(
t\right) \right) $ is the dissipator of the master equation through which
the relevant decaying processes such as the spontaneous emission and
absorption can be specified. Essentially, the medium is envisaged as a
single, huge and stationary, 3-level \textquotedblleft
atom\textquotedblright , so that only the internal atomic degrees of freedom
are considered in the last equation. To include the effects of
nonrelativistic thermal random motion of atoms, a plausible way is to add a
Doppler energy term $\hbar \mathbf{k}\cdot \mathbf{v}$ directly to the
Hamiltonian in eq.($\ref{Linblad}$), where $\mathbf{k}$ is the wave vector of
the applied laser field, and $\mathbf{v}$ is the relative velocity between
the atom and the light source. On account of the thermal motion of atoms, $%
\mathbf{v}$ is randomized, and therefore the final solution of eq.($\ref%
{Linblad}$) has to be determined \emph{statistically}. Denoting $\hat{\rho}%
^{\left( \mathbf{v}\right) }$ as the solution of eq.($\ref{Linblad}$) with a
given $\mathbf{v}$, then the desired density matrix can be obtained by
taking average over all possible velocities which are typically described by
Maxwell-Boltzmann distribution when the gas is at thermal equilibrium. It is
expected that such an averaging process smears the atomic coherence and thus
contributes to the overall effect of decoherence\emph{\ }\cite{Javan} .

We note that, for taking the effect of Doppler shift into account the above
formula is valid only when the \textquotedblleft atom\textquotedblright\ is
stationary or, equivalently, when the frame of reference is fixed on the
\textquotedblleft atom\textquotedblright . As we aim to obtain the
statistically averaged density matrix over an ensemble of identical EIT
systems moving with different velocities, a particular frame of reference
must be specified in which each $\hat{\rho}^{\left( \mathbf{v}\right) }$ can
be solved and properly\ weighted on the common ground. This would involve
the transformation of motion equations under the principle of Galilean
relativity. However, in the presence of driving fields, the atoms are by no
means in constant motion as they do exchange momenta with the light fields
by absorbing or emitting photons from time to time. Furthermore, the atoms
are liable to couple to vacuum via spontaneous emission of photons to all
directions, giving rise to the random recoil of the atoms. These
circumstances suggests that the moving atoms do not serve as an ideal frame
of reference in our scheme. Alternatively, we may seek to solve the matrix
elements of all $\hat{\rho}^{\left( \mathbf{v}\right) }\left( t\right) $ and
carry out the ensemble average in the laboratory frame \cite{Bo Zhao} . In
doing so, we might have to consider the sample gas as a lump of medium
rather than a single \textquotedblleft atom\textquotedblright . Accordingly,
it is more convenient to work in the framework of Schr\"{o}dinger equation
instead of the master equation. There are two reasons: the gauge invariance
of Schr\"{o}dinger equation under the Galilean transformation are well
understood. Second, the kinetic energy can be naturally included in the Schr%
\"{o}dinger equation, so that the motional coupling, namely, the coupling
between the external degrees of freedom and the internal degrees of freedom,
can be restored and properly dealt with, although they are normally
overwhelmed by the light-atom interaction. The details of the derivation of
the motion equations will be described later.

The organization of this paper is as follows. In Sec. II, we derive the
general forms of the motion equations of the atomic medium under arbitrary
Galilean transformation boosted along the $z$-direction. The numerical
results are presented in Sec. III. Comparison with experimental data are
made. Finally, some concluding remarks are given in Sec. IV.

\section{Formalism}

We consider a medium consisting of $\Lambda $-type 3-level atoms with two
metastable ground states as shown in Fig. 1. The atoms in this medium are
all noninteracting and excited by two laser fields. The probe pulse, which
drives the $\left\vert 1\right\rangle \leftrightarrow $ $\left\vert
3\right\rangle $ transition is characterized by a central frequency $\omega
_{p}$ and a wave vector $\mathbf{k}_{p}$. On the other hand, the $\left\vert
2\right\rangle \leftrightarrow $ $\left\vert 3\right\rangle $ transition is
driven by another laser field with frequency $\omega _{c}$ and wave vector $%
\mathbf{k}_{c}$. For all practical purposes, the probe and couple fields can
be applied with different relative orientation, depending on the
experimental applications. For simplicity, we shall assume in the following
derivations that the probe field propagates along the $z$ direction and the
coupling field propagates with an angle $\theta $ with respect to the $z$%
-axis and the 3-level atoms are in a cigar-shaped trap which can be
considered as an one dimensional system. Now the Hamiltonian of the system
can be given by

\begin{figure}[htbp]\begin{center}
\includegraphics[width=1.6734in]{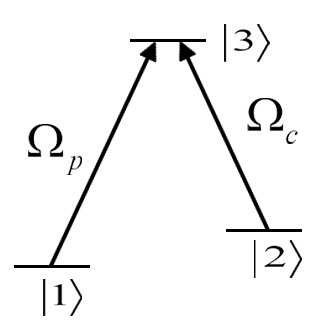}
\caption{(Color online) The energy levels of a 3-level atom of $\Lambda $-type. The transitions $
\left\vert 1\right\rangle \leftrightarrow $ $\left\vert 3\right\rangle $ and
$\left\vert 2\right\rangle \leftrightarrow $ $\left\vert 3\right\rangle $
are driven by laser fields with Rabi frequencies $\Omega _{p}$ and $\Omega
_{c}$ correspondingly.}
\label{sag}
\end{center}\end{figure}

\begin{equation}
H=H_{0}+H_{I},
\end{equation}%
where

\begin{equation}
H_{0}=\left[
\begin{array}{ccc}
\frac{{p}^{2}}{2m}+E_{1} & 0 & 0 \\
0 & \frac{{p}^{2}}{2m}+E_{2} & 0 \\
0 & 0 & \frac{{p}^{2}}{2m}+E_{3}%
\end{array}%
\right] ,  \label{H0}
\end{equation}%
is the unperturbed part and

\begin{eqnarray}
&&H_{I} \\
&=&\frac{\hbar }{2}\left[
\begin{array}{ccc}
0 & 0 & \Omega _{p}^{\ast }e^{-i\left( k_{p}z-\omega _{p}t\right) } \\
0 & 0 & \Omega _{c}^{\ast }e^{-i\left( \tilde{k}_{c}z-\omega _{c}t\right) }
\\
\Omega _{p}e^{i\left( k_{p}z-\omega _{p}t\right) } & \Omega _{c}e^{i\left(
\tilde{k}_{c}z-\omega _{c}t\right) } & 0%
\end{array}%
\right] ,  \nonumber  \label{Hi}
\end{eqnarray}%
describe the interaction between EM field and the atom in the dipole
approximation and rotating-wave approximation. Here, $E_{j}$ is the energy
of the electronic level $\left\vert j\right\rangle $, $\tilde{k}%
_{c}=k_{c}\cos \theta $ and $\Omega _{c}$ and $\Omega _{p}$ denote the Rabi
frequencies for the transitions $\left\vert 2\right\rangle \leftrightarrow $
$\left\vert 3\right\rangle $ and $\left\vert 1\right\rangle \leftrightarrow $
$\left\vert 3\right\rangle $ respectively. In the current problem, we assume
that $\Omega _{p}=\Omega _{p}\left( z,t\right) $ and $\Omega _{c}=\Omega
_{c}\left( t\right) $.

In the continuum limit, a 3-level atomic medium can be described by a
3-component spinor field,
\begin{equation}
\mathbf{\Psi }\left( \mathbf{r},t\right) =\left(
\begin{array}{c}
\psi _{1}\left( \mathbf{r},t\right) \\
\psi _{2}\left( \mathbf{r},t\right) \\
\psi _{3}\left( \mathbf{r},t\right)%
\end{array}%
\right) ,
\end{equation}%
where $\psi _{j}(\mathbf{r},t)$ is the atomic field operator which
annihilates an atom in the internal state $\left\vert j\right\rangle $ that
is positioned at $z$. In terms of the atomic field operators, the energy
functional of the above EIT system is given by
\begin{equation}
\mathcal{E}\left[ \mathbf{\Psi }^{\mathbf{\dag }},\mathbf{\Psi }\right]
=\int d^{3}r\mathbf{\Psi }^{\mathbf{\dag }}\left( \mathbf{r},t\right) H%
\mathbf{\Psi }\left( \mathbf{r},t\right)  \label{energy functional}
\end{equation}%
The motion equations for all $\psi _{j}(\mathbf{r},t)$ can be derived from
the Hartree variational principle, namely,
\begin{equation}
i\hbar \frac{\partial \psi _{j}}{\partial t}=\frac{\delta \mathcal{E}\left[
\mathbf{\Psi }^{\mathbf{\dag }},\mathbf{\Psi }\right] }{\delta \psi
_{j}^{\ast }},\qquad \left( j=1,2,3\right)  \label{Hartree}
\end{equation}%
and consequently, we obtain

\begin{equation}
i\hbar \frac{\partial \psi _{1}}{\partial t}=\left[ -\frac{\hbar ^{2}}{2m}%
\frac{\partial ^{2}}{\partial z^{2}}+E_{1}\right] \psi _{1}+\frac{\hbar }{2}%
\Omega _{p}^{\ast }e^{-i\left( k_{p}z-\omega _{p}t\right) }\psi _{3},
\label{SGE1 v=0}
\end{equation}

\begin{equation}
i\hbar \frac{\partial \psi _{2}}{\partial t}=\left[ -\frac{\hbar ^{2}}{2m}%
\frac{\partial ^{2}}{\partial z^{2}}+E_{2}\right] \psi _{2}+\frac{\hbar }{2}%
\Omega _{c}^{\ast }e^{-i\left( \tilde{k}_{c}z-\omega _{c}t\right) }\psi _{3},
\label{SGE2 v=0}
\end{equation}

\begin{eqnarray}
&&i\hbar \frac{\partial \psi _{3}}{\partial t}=\left[ -\frac{\hbar ^{2}}{2m}%
\frac{\partial ^{2}}{\partial z^{2}}+E_{3}\right] \psi _{3}+\frac{\hbar }{2}%
\Omega _{p}e^{i\left( k_{p}z-\omega _{p}t\right) }\psi _{1}  \nonumber \\
&&+\frac{\hbar }{2}\Omega _{c}e^{i\left( \tilde{k}_{c}z-\omega _{c}t\right)
}\psi _{2},  \label{SGE3 v=0}
\end{eqnarray}%
Here we have ignored the transverse motion of the atoms in the $xy$-plane
since the Gaussian probe pulse propagates along the $z$-direction only.

It is convenient to describe the atomic properties by means of the local
density operator $\hat{\rho}$ whose matrix elements are defined as bilinear
products of atomic fields, i.e., $\rho _{ij}=\psi _{i}\psi _{j}^{\ast }$.
Accordingly, the $i$-th diagonal matrix element corresponds to the density
of atoms in the state $\left\vert j\right\rangle $. In the absence of any
decaying process, the total density may be normalized to $\int
dz\sum_{j}\left\vert \psi _{j}\right\vert ^{2}=NL$, where $N$ is total
particle number and $L$ is the length of the system. The quantum coherence
between the states $\left\vert i\right\rangle $ and $\left\vert
j\right\rangle $ is represented by the off-diagonal matrix element $\rho
_{ij}$. In particular, the matrix element, $\rho _{12}$, is of central
importance in the current problem, which dominates the dynamics of the
storage and retrieval of the probe light pulse. Additionally, the coherence
between $\left\vert 1\right\rangle $ and $\left\vert 3\right\rangle $
determines the propagation of the probe field inside the medium. In the
slowly varying envelope approximation \cite{M. Scully} , the dynamics of the
probe field is governed by the Maxwell-Schr\"{o}dinger equation

\begin{equation}
\left( \frac{\partial }{\partial z}+\frac{1}{c}\frac{\partial }{\partial t}%
\right) \Omega _{p}=-i\eta \rho _{31},  \label{Maxwe-Schroedinger}
\end{equation}%
where $\eta =3\lambda _{L}^{2}\mathcal{N}a\Gamma /4\pi $ with $\mathcal{N}$
being the number per unit length of the medium), $\Gamma $ the spontaneous
decay rate of $\left\vert 3\right\rangle $, $a$ the branch ratio of the
decay from $\left\vert 3\right\rangle $ to $\left\vert 1\right\rangle $, and
$\lambda _{L}$ being the wavelength of the laser field.

Note that eqs.($\ref{SGE1 v=0}$)-($\ref{SGE3 v=0}$) can be interpreted as the
governing equations of a moving continuous medium seen by a co-moving
observer, although they take the form of coupled Schr\"{o}dinger equations.
In the nonrelativistic quantum mechanics, the Galilean relativity ensures
that the Schr\"{o}dinger equation is form-invariant under the Galilean
transformation, $\mathbf{r}\rightarrow \mathbf{r}-\mathbf{v}t$, $%
t\rightarrow t$, for any two frames of reference that are in relative
uniform translational motion with a velocity $\mathbf{v}$ \cite{E.
Merzbacher} . To be specific, let us assume that the boosted system moves
with a velocity $\mathbf{v}=v\mathbf{\hat{z}}$ relative to the laboratory
frame. Accordingly, the Galilean principle of relativity demands the gauge
dependence of two different but equivalent quantum mechanical states
\begin{equation}
\mathbf{\Psi }\left( z,t\right) =\mathbf{\Psi }^{\left( v\right) }\left(
z^{\prime },t^{\prime }\right) e^{-i\left( mv^{2}t/2-mvz\right) /\hbar },
\label{Galilean transform}
\end{equation}%
where $\left( z,t\right) $ and $\left( z^{\prime },t^{\prime }\right) $
denote the space-time coordinates in the laboratory frame and the boosted
frame respectively, which are transformed by $z^{\prime }=z-vt$, $t^{\prime
}=t$. Moreover, we require that the pulse profiles in different frames are
related by,%
\begin{equation}
\Omega _{p}\left( z,t\right) =\Omega _{p}^{\prime }\left( z^{\prime
},t^{\prime }\right) =\Omega _{p}^{\prime }\left( z-vt,t\right) ,
\label{field profile}
\end{equation}%
and thus eqs.(\ref{SGE1 v=0})-(\ref{SGE3 v=0}) become

\begin{eqnarray}
&&i\hbar \frac{\partial }{\partial t}\psi _{1}^{\left( v\right) }\left(
z-vt,t\right)  \label{wave lab  GF1} \\
&=&\left[ -\frac{\hbar ^{2}}{2m}\frac{\partial ^{2}}{\partial z^{2}}-iv\hbar
\frac{\partial }{\partial z}+\hbar \omega _{1}\right] \psi _{1}^{\left(
v\right) }\left( z-vt,t\right)  \nonumber \\
&+&\frac{\hbar }{2}\Omega _{p}^{\ast }\left( z,t\right) e^{-i\left(
k_{p}z-\omega _{p}t\right) }\psi _{3}^{\left( v\right) }\left( z-vt,t\right)
,  \nonumber
\end{eqnarray}

\begin{eqnarray}
&&i\hbar \frac{\partial \psi _{2}^{\left( v\right) }\left( z-vt,t\right) }{%
\partial t}  \label{wave lab GF2} \\
&=&\left[ -\frac{\hbar ^{2}}{2m}\frac{\partial ^{2}}{\partial z^{2}}-iv\hbar
\frac{\partial }{\partial z}+\hbar \omega _{2}\right] \psi _{2}^{\left(
v\right) }\left( z-vt,t\right)  \nonumber \\
&+&\frac{\hbar }{2}\Omega _{c}^{\ast }\left( t\right) e^{-i\left( \tilde{k}%
_{c}z-\omega _{c}t\right) }\psi _{3}^{\left( v\right) }\left( z-vt,t\right) ,
\nonumber
\end{eqnarray}

\begin{eqnarray}
&&i\hbar \frac{\partial \psi _{3}^{\left( v\right) }\left( z-vt,t\right) }{%
\partial t}  \label{wave lab GF3} \\
&=&\left[ -\frac{\hbar ^{2}}{2m}\frac{\partial ^{2}}{\partial z^{2}}-iv\hbar
\frac{\partial }{\partial z}+\hbar \omega _{3}\right] \psi _{3}^{\left(
v\right) }\left( z-vt,t\right)  \nonumber \\
&+&\frac{\hbar }{2}\Omega _{p}\left( z,t\right) e^{i\left( k_{p}z-\omega
_{p}t\right) }\psi _{1}^{\left( v\right) }\left( z-vt,t\right)  \nonumber \\
&+&\frac{\hbar }{2}\Omega _{c}\left( t\right) e^{i\left( \tilde{k}%
_{c}z-\omega _{c}t\right) }\psi _{2}^{\left( v\right) }\left( z-vt,t\right) ,
\nonumber
\end{eqnarray}%
where $\psi _{j}^{\left( v\right) }$ ($j=1,2,3$) denote the\ atomic fields
in the medium boosted with velocity $v$ and $\omega _{j}=E_{j}/\hbar $.

To solve eqs.($\ref{wave lab GF1}$)-($\ref{wave lab GF3}$) in a more efficient
manner (less grid points and higher accuracy), we extract a fast oscillating
phase (both spatially and temporarily) from each $\psi _{j}^{\left( v\right)
}$ by writing

\begin{equation}
\left(
\begin{array}{c}
\psi _{1}^{\left( v\right) } \\
\psi _{2}^{\left( v\right) } \\
\psi _{3}^{\left( v\right) }%
\end{array}%
\right) =\left(
\begin{array}{c}
\phi _{1}^{\left( v\right) }\left( z,t\right) e^{-i\omega _{1}t} \\
\phi _{2}^{\left( v\right) }\left( z,t\right) e^{-i\left( \omega _{p}-\omega
_{c}+\omega _{1}\right) t+i(k_{p}-\tilde{k}_{c})z} \\
\phi _{3}^{\left( v\right) }\left( z,t\right) e^{-i\left( \omega _{p}+\omega
_{1}\right) t+ik_{p}z}%
\end{array}%
\right) ,  \label{wave fn. extract}
\end{equation}%
where $\phi _{j}^{\left( v\right) }$ represents the slowly varying part for $%
\psi _{j}^{\left( v\right) }$. Substituting eq.($\ref{wave fn. extract}$) into
eqs.($\ref{wave lab GF1}$)-($\ref{wave lab GF3}$) and taking the spontaneous
decay of the excited level $\left\vert 3\right\rangle $ into account, we
then obtain the following motion equations for $\phi _{j}^{\left( v\right) }$
in the laboratory frame,

\begin{eqnarray}
&&i\hbar \frac{\partial \phi _{1}^{\left( v\right) }\left( z,t\right) }{%
\partial t}  \label{wave slowly varying 1} \\
&=&\left[ -\frac{\hbar ^{2}}{2m}\frac{\partial ^{2}}{\partial z^{2}}-iv\hbar
\frac{\partial }{\partial z}\right] \phi _{1}^{\left( v\right) }\left(
z,t\right) +\frac{\hbar }{2}\Omega _{p}^{\ast }\phi _{3}^{\left( v\right)
}\left( z,t\right)  \nonumber
\end{eqnarray}

\begin{eqnarray}
&&i\hbar \frac{\partial \phi _{2}^{\left( v\right) }\left( z,t\right) }{%
\partial t}  \label{wave slowly varying 2} \\
&=&\left[ -\frac{\hbar ^{2}}{2m}\frac{\partial ^{2}}{\partial z^{2}}-iv\hbar
\frac{\partial }{\partial z}-\frac{i\left( k_{p}-\tilde{k}_{c}\right) \hbar
^{2}}{m}\frac{\partial }{\partial z}\right] \phi _{2}^{\left( v\right)
}\left( z,t\right)  \nonumber \\
&+&\hbar \left( \triangle _{p}-\triangle _{c}+\triangle _{v}\right) \phi
_{2}^{\left( v\right) }\left( z,t\right) +\frac{\hbar }{2}\Omega _{c}^{\ast
}\phi _{3}^{\left( v\right) }\left( z,t\right) ,  \nonumber
\end{eqnarray}

\begin{eqnarray}
&&i\hbar \frac{\partial \phi _{3}^{\left( v\right) }\left( z,t\right) }{%
\partial t}  \label{wave slowly varying 3} \\
&=&\left[ -\frac{\hbar ^{2}}{2m}\frac{\partial ^{2}}{\partial z^{2}}-iv\hbar
\frac{\partial }{\partial z}-\frac{ik_{p}\hbar ^{2}}{m}\frac{\partial }{%
\partial z}\right] \phi _{3}^{\left( v\right) }\left( z,t\right)  \nonumber
\\
&+&\left[ \hbar \left( \triangle _{p}+k_{p}v\right) -i\frac{\Gamma }{2}+%
\frac{\hbar ^{2}k_{p}^{2}}{2m}\right] \phi _{3}^{\left( v\right) }\left(
z,t\right)  \nonumber \\
&+&\frac{\hbar }{2}\Omega _{p}\phi _{1}^{\left( v\right) }\left( z,t\right) +%
\frac{\hbar }{2}\Omega _{c}\phi _{2}^{\left( v\right) }\left( z,t\right) ,
\nonumber
\end{eqnarray}%
where $\triangle _{p}=\omega _{3}-\omega _{1}-\omega _{p},$ $\triangle _{c}=$
$\omega _{3}-\omega _{2}-\omega _{c}$ are the detunings of probe and
coupling lasers respectively, and $\triangle _{v}=\left( k_{p}-\tilde{k}%
_{c}\right) v+\hbar (k_{p}-\tilde{k}_{c})^{2}/2m$ denotes the motion-induced
frequency shift. In eq.($\ref{wave slowly varying 3}$), the spontaneous decay
of the excited level $\left\vert 3\right\rangle $ occurring with a rate $%
\Gamma $ is included phenomenologically. Note that the term, $k_{p}v$ and $%
\tilde{k}_{c}v$ entering the right-hand-side of eq.($\ref{wave slowly varying
2}$) and eq.($\ref{wave slowly varying 3}$), are the nonrelativistic Doppler
shift which turn out to be the detunings for the probe and coupling laser
respectively in the laboratory frame. In addition, it is noteworthy to point
out that $\hbar (k_{p}-\tilde{k}_{c})^{2}/2m$ \ represents the frequency
shift caused by the recoil of atom.

The dephasing owing to the thermal motion of atoms is incorporated in the
system dynamics by replacing $\rho _{31}$ with $\left\langle \rho
_{31}\right\rangle _{T}$ in the right-hand-side of eq.($\ref%
{Maxwe-Schroedinger}$), namely,
\begin{equation}
\left( \frac{\partial }{\partial z}+\frac{1}{c}\frac{\partial }{\partial t}%
\right) \Omega _{p}=-i\eta \left\langle \rho _{31}\left( z,t\right)
\right\rangle _{T},  \label{Maxwell eq. probe}
\end{equation}%
where
\begin{eqnarray*}
&&\left\langle \rho _{31}\left( z,t\right) \right\rangle _{T} \\
&=&\sqrt{\frac{m}{2k_{B}T\pi }}\int dv\phi _{3}^{\left( v\right) }\left(
z,t\right) \phi _{1}^{\left( v\right) \ast }\left( z,t\right)
e^{-mv^{2}/2k_{B}T},
\end{eqnarray*}%
the thermally averaged atomic coherence between $\left\vert 1\right\rangle $
and $\left\vert 3\right\rangle $, which is taken over the Maxwell-Boltzmann
distribution of velocity at a given temperature $T$.

Finally, we give a brief account for the numerical method which we have
employed to integrate eqs.($\ref{wave slowly varying 1}$)-($\ref{Maxwell eq.
probe}$). Since we have used very high resolution to resolve the fine
structures of the wave functions and the light pulse during their time
evolutions, the commonly used second-order Crank-Nicolson method turns out
to be inefficient in the current problem despite that it is unconditionally
stable. Owing to its implicitness in time, the Crank-Nicolson method would
require an exceedingly large resultant matrix for the four coupled
equations, eqs.($\ref{wave slowly varying 1}$)-($\ref{Maxwell eq. probe}$), when
a high spatial resolution is demanded. With this regard, explicit methods
are more efficient for the current problem. Here we use the method of lines
with spatial discretization by highly accurate Fourier pseudospectral method
and time integration by adaptive Runge-Kutta method of orders 2 and 3
(RK23), such that the accuracy in time is at least of second order, and the
accuracy in space is of exponential order. In the following numerical
computations, we will first determine $\phi _{j}^{\left( v\right) }\left(
z,t\right) $ by numerical integration. Having obtained all $\phi
_{j}^{\left( v\right) }\left( z,t\right) $, we then calculate $\left\langle
\rho _{31}\left( z,t\right) \right\rangle _{T}$ to determine the profile of $%
\Omega _{p}$ at the instant $t$.

\section{Results and discussions}

In this section, we apply the aforementioned scheme to simulate the effect
of Doppler broadening for both SL and LS, and compare the numerical results
with the experimental data. Before we present the results, some remarks are
given. To begin with, let us consider the limiting case with $k_{p}=k_{c}$
and $\theta =0$, where the shift $\triangle _{v}$ in the right-hand-side of
eq.($\ref{wave slowly varying 2}$) can be exactly cancelled out, indicating
that the EIT dynamics can be robust against the effect of Doppler broadening
when probe and couple are co-propagating. However, we note that in most
experiments, $k_{p}$ differs from $k_{c}$ in the transition scheme, or a
small $\theta $ is required to separate the weak probe and strong coupling
fields in the detection scheme. This suggests that the effects of Doppler
broadening is inevitable in practical applications, despite that they are
essentially small in the co-propagating situation. On the contrary, setting $%
k_{p}=k_{c}$ and $\theta =\pi $, $\triangle _{v}$ is maximized for all $v$
and can not be eliminated in any case and thus a substantial loss of
ground-state coherence is expected in the counter-propagating geometry.
Regarding the fact that the decoherence is much pronounced in the latter
case, plus that the counter-propagating EIT is the central mechanism for
generating stationary light pulses \cite{Lin} , here we restrict our
attention on the counter-propagating EIT. As a matter of fact, in our
co-propagating EIT experiment in which the effect of the thermal motion of
the cold atoms is minimized, the decoherence rate is found to be less than $%
10^{-3}\Gamma $ ($\Gamma =2\pi \times 5.9$ MHz), indicating that the
relaxations caused by some intrinsic effects, such as the stray magnetic
field, collisions, and linewidth and frequency fluctuation of the laser
fields, etc., are very small which add up to give a decoherence rate less
than\textbf{\ }$10^{-3}\Gamma $\textbf{.} Such a small decoherence rate
suggests that the decay of the output probe fields in our
counter-propagating experiment are mainly from the thermal motion. In fact,
adding such small decoherence rate in our calculation changes the numerical
outcome very little and therefore we shall not consider the intrinsic
decoherence in the numerical simulations.

The comparisons between experimental data and numerical results of SL are
made in Fig.2 for three different experiments. We carried out the
measurements in a cigar-shaped cloud of laser-cooled Rb atoms \cite{Chou} in
which the probe pulse and coupling field are counter-propagating along the
major axis of the cloud. The Rabi frequency $\Omega _{c}$ and the optical
density $OD\left( =2L\eta /\Gamma \right) $ for the three samples are
estimated to be $\Omega _{c}=0.85\Gamma $, $OD=30$, $\Omega _{c}=0.69\Gamma $%
, $OD=41$ and $\Omega _{c}=0.77\Gamma $, $OD=48$ respectively. Both $\Omega
_{c}$ and $OD$ are estimated by the method described in \cite{Lin} and have
an uncertainty of about $\pm 5\%.$ The probe pulses used in the experiment
are sufficiently weak such that the corresponding Rabi frequency in the
calculation does not affect the prediction of the output probe pulse. In the
numerical simulations with $128$ velocity groups, we set $\Omega
_{c}=0.825\Gamma $, $OD=30$, $T=290$ $\mu $K, $\Omega _{c}=0.665\Gamma $%
\textbf{, }$OD=41.3$\textbf{, }$T=305$\textbf{\ }$\mu $K and $\Omega
_{c}=0.75\Gamma $\textbf{, }$OD=48$\textbf{, }$T=280$\textbf{\ }$\mu $K\ to
get the best fit for the three experiments of SL shown in Fig.2(a)-2(c),
respectively. It should be mentioned that the temperature of the SL
experiment in Fig.2(a)\textbf{\ }has been determined by another different
numerical method to give the value of $T=296$ $\mu $K \cite{Wu} , which is
very close to our prediction. The experimental data and numerical results of
storage and retrieval of a light pulse are shown in Fig.3. In the
simulations of Fig.3, we only adjust the temperature to $T=240$ $\mu $K
while keep the values of $\Omega _{c}$ and $OD$ in Fig.2(a) unchanged to get
the best fit. The discrepancies in the temperatures obtained from the
numerical simulations in Fig.2(a) and Fig.3 are very reasonable as compared
with the expected temperature and fluctuations of the laser-cooled Rb atoms,
yet $\Omega _{c}$ and $OD$ so obtained are in good agreement with the
experimental parameters. Therefore, we have quantitatively demonstrated the
validity and accuracy of the numerical method presented in this work. In
addition, we find this method can be used to determine the temperature along
the major axis of the cigar-shaped atom cloud which we were not able to
measure previously.

The influence of the thermal effect is readily revealed by the considerably
diminished intensity of the output probe beam. In Fig.4(a), the intensity
profiles of the output probe pulse at various temperatures are shown for SL.
As expected, the output intensity decreases when the temperature increases.
A simple explanation for this is that the higher the temperature, the
broader the velocity distribution is. Thus the effective two-photon detuning
(the averaged Doppler shift) becomes larger and the output probe pulse is
significantly suppressed. Fig.4(b) shows that the peak of the output probe
decays exponentially as a function of temperature.

\begin{figure}[htbp]\begin{center}
\includegraphics[width=3.2794in]{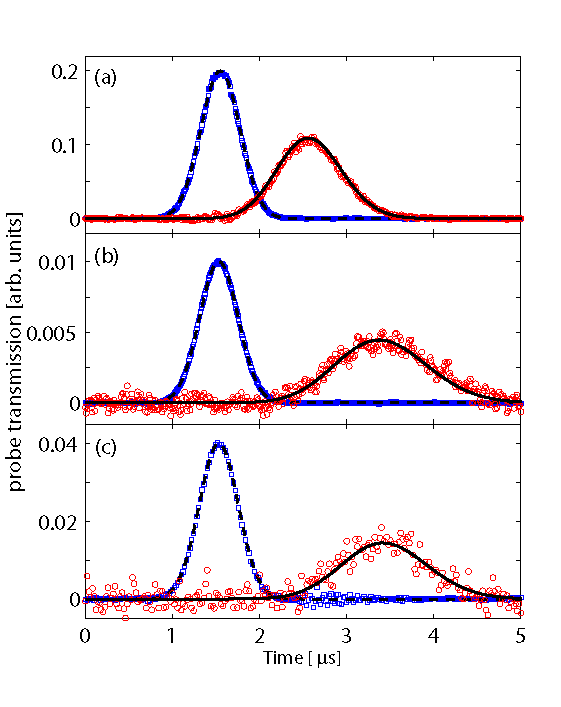}
\caption{(Color online) The best fitting
of the numerical simulation with (a) $\Omega _{c}=0.825\Gamma $, $OD=30$, $%
T=290$ $\protect\mu $K; (b) $\Omega _{c}=0.665\Gamma $, $OD=41.3$, $T=305$ $%
\protect\mu $K; (c) $\Omega _{c}=0.75\Gamma $, $OD=48$, $T=280$ $\protect\mu
$K for SL. The squares and circles represent the intensities of the
experimentally measured input probe pulse and output probe pulse,
respectively. The input probe pulse is scaled down by a factor of $0.2$ in
(a), $0.01$ in (b) and $0.04$ in (c), respectively. The dashed and solid
lines represent the nurmerical simulations for the intensities of the input
probe pulse and the output probe pulse, respectively.}
\label{fig2}
\end{center}\end{figure}

\begin{figure}[htbp]\begin{center}
\includegraphics[width=3.6365in]{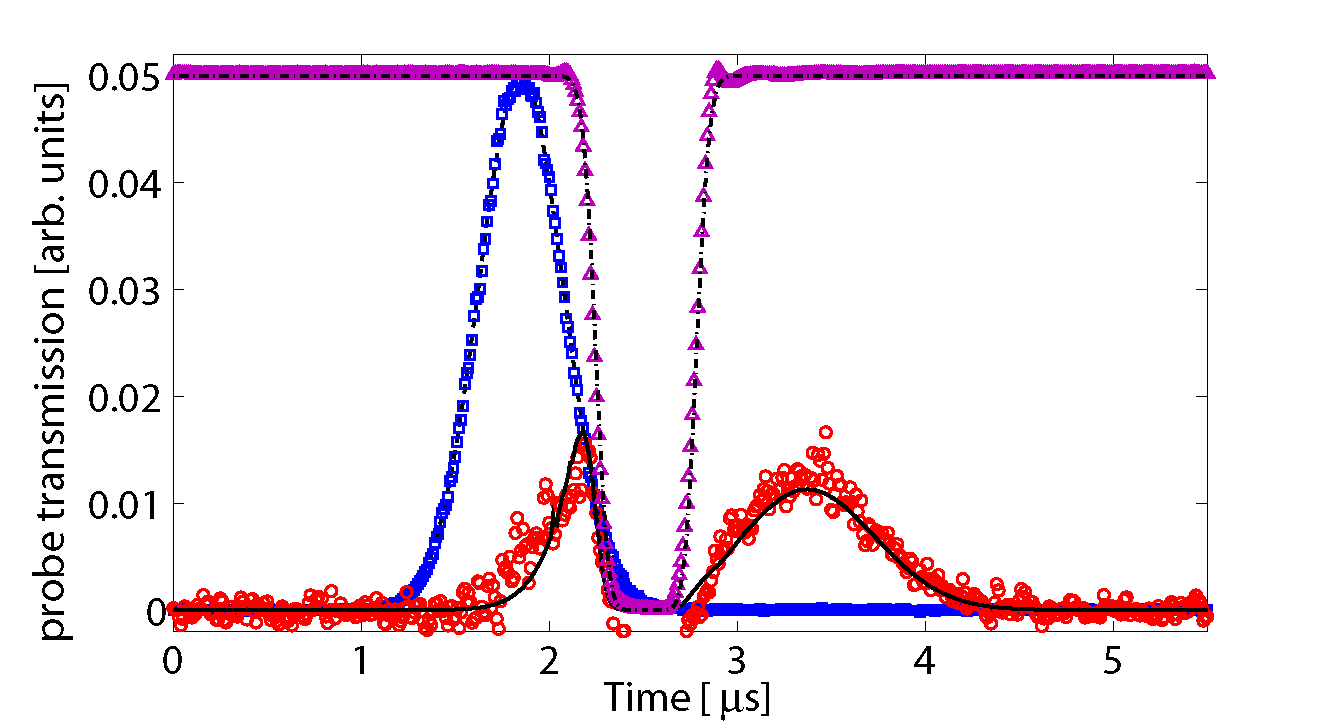}
\caption{(Color online) The best fitting
of the numerical simulation with $\Omega _{c}=$ $0.825\Gamma $, $OD=30$, $%
T=240$ $\protect\mu $K for LS. The squares and circles are the
experimentally measured intensities of the input probe pulse and output
probe pulse, respectively. The dashed line and solid line represent the the
nurmerical simulations for the intensities of the input probe pulse and
output probe pulse, respectively. The input probe pulse is scaled down by a
factor of $0.05$.}
\label{fig3}
\end{center}\end{figure}

For all practical purposes, it is more instructive to examine the
ground-state coherence, $\left\langle \rho _{21}\right\rangle _{T}$ , rather
than the output intensity of the probe beam, since there are no light fields
during the process of storage, and once the stored pulse is retrieved, it
restores SL again. Although $\left\langle \rho _{21}\right\rangle _{T}$ can
not be measured directly, its magnitude determines the intensity of the
retrieved probe pulse when the coupling field is turned off. Very recently,
in analyzing the feasibility of measuring the ground-state coherence in an
EIT, Zhao \textit{et al}. \cite{Bo Zhao} suggested that $\left\langle \rho
_{21}\right\rangle _{T}$ decays like a Gaussian function during LS. On this
basis, it can be further shown that
\begin{equation}
\left\langle \rho _{21}\left( z,t\right) \right\rangle _{T}=\left\langle
\rho _{21}(z,t=0)\right\rangle _{T}e^{-(kv_{s})^{2}t^{2}}e^{2ikz},
\label{rho21 decay}
\end{equation}%
where $v_{s}=\sqrt{2k_{B}T/m}$ is the one-dimensional root-mean-square
velocity and $k=c^{-1}(\omega _{3}-\omega _{1})$. Now let us consider the
case of the storage time of $1.4$ $\mu $s for various temperatures $%
T=100,300,500$ $\mu $K. In the calculation, the probe pulse enters the
medium under $\Omega _{c}=0.75\Gamma $ and $OD=32$ both of which do not
affect the decay behavior during the storage; the input probe pulse and the
timing of switching off the coupling field are the same as those shown in
Fig.3. For such a storage time and atom temperatures the atomic thermal
motion is expected to well smear the quantum memory of the probe pulse which
is stored as the ground-state coherence. Because eq.($\ref{rho21 decay}$) is a
function of $z$ and $t$ and the time-dependence only comes from the Gaussian
function, to eliminate the $z$ dependence of the ground-state coherence, we
plot the function $R_{21}\left( t,T\right) =\left\vert \int \left\langle
\rho _{21}\left( z,t\right) \right\rangle _{T}dz\right\vert $ in Fig.5. We
see that the $R_{21}\left( t,T\right) $ indeed decays Gaussian-like with
time as predicted in \cite{Bo Zhao} . Now let us denote $\tau $ as the $1/e$
width of the Gaussian function. Accordingly, $\tau $ of the fitted Gaussian
function of the ground-state coherence in the simulation is found to be
close to the value predicted by eq.($\ref{rho21 decay}$). Furthermore, we have
verified that $\tau \propto T^{-1/2}$, and the close agreement with the
theoretical predictions of eq.($\ref{rho21 decay}$) are also shown in Fig.5.
Since the thermal motion would randomize the spatial profile of the
ground-state coherence during the storage, the intensity of the retrieved
probe pulse is thus much smaller than the stored one as shown in Fig.3.

The ground-state coherence for SL can be studied in a similar manner. It is
expected that the ground-state coherence decays more slowly in SL than in
LS, since the continuous optical excitations from the coupling field can
retard the loss of atomic coherence led by the randomization of atoms'
thermal motion. We simulate the process of SL with $\Omega _{c}=$ $%
0.75\Gamma $, and $OD=400$ at various temperatures. Here we have chosen a
very large optical density $OD=400$ to ensure that the probe pulse can stay
in the medium for a sufficiently long period. We plot $R_{21}\left(
t,T\right) $ as a function of time for various temperatures after the probe
pulse has entirely merged into the medium, and the numerical results suggest
that $R_{21}\left( t,T\right) $ decays exponentially with a rate $\kappa $.
In the following, we apply some previous theoretical results based on the
steady-state solution of the optical-Bloch equation to derive an analytic
estimate of $\kappa $. For simplicity, we shall assume that the EIT
transparency bandwidth is much larger than the frequency bandwidth of the
probe pulse, such that the decay of the probe pulse or the ground-state
coherence is mainly determined by the absorption in the center of the
transparency window of the EIT spectrum. Because all decoherence mechanisms
other than the atomic thermal motion are neglected, the steady-state
solution of $\rho _{31}$ of a rest EIT medium is given by \cite{Chen}

\begin{figure}[htbp]\begin{center}
\includegraphics[width=3.2154in]{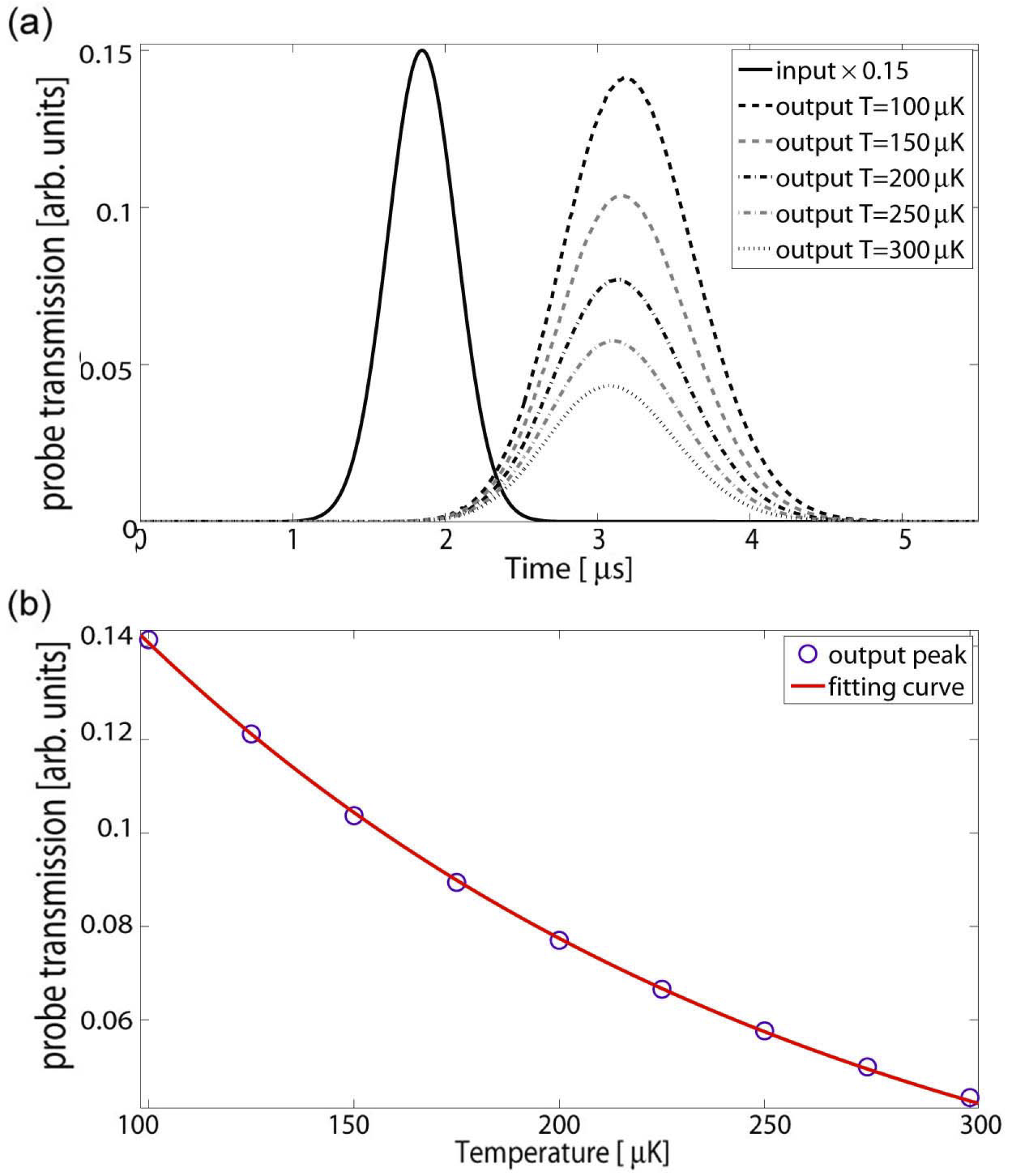}
\caption{(Color online) The best fitting
of the numerical simulation with $\Omega _{c}=$ $0.825\Gamma $, $OD=30$, $%
T=240$ $\protect\mu $K for LS. The squares and circles are the
experimentally measured intensities of the input probe pulse and output
probe pulse, respectively. The dashed line and solid line represent the the
nurmerical simulations for the intensities of the input probe pulse and
output probe pulse, respectively. The input probe pulse is scaled down by a
factor of $0.05$.}
\label{fig3}
\end{center}\end{figure}

\begin{equation}
\frac{\rho _{31}}{\Omega _{p}}=\frac{\triangle _{c}-\triangle _{p}}{\Omega
_{c}^{2}/2+2[\triangle _{p}-i\Gamma /2][\triangle _{c}-\triangle _{p}]}.
\label{rho31}
\end{equation}%
Given that the probe and coupling fields are both resonant with the atomic
transition frequencies in the laboratory frame, thus for an atom moving with
a velocity $v$, we have $\triangle _{p}=kv$ and $\triangle _{c}=-kv$ in the
above equation. In the presence of a strong coupling field, $\Omega _{c}\gg
kv_{s}$, the imaginary part of eq.($\ref{rho31}$) can be approximated by
\begin{equation}
\textit{Im}\left[ \frac{\rho _{31}(v)}{\Omega _{p}}\right] \approx \frac{%
4k^{2}v^{2}\Gamma }{\Omega _{c}^{4}},  \label{im-rho31}
\end{equation}%
which is related to the absorption coefficient $\alpha $ of a
Doppler-broadened medium by

\begin{equation}
\alpha =\eta \left\langle \textit{Im}\left[ \frac{\rho _{31}(v)}{\Omega _{p}}%
\right] \right\rangle _{T}=\frac{4\eta k^{2}v_{s}^{2}\Gamma }{\Omega _{c}^{4}%
}.  \label{absorption}
\end{equation}%
The absorption gives rise to the attenuation of the propagating light pulse,
that is described by the Beer's law, namely,
\begin{equation}
\frac{R_{21}\left( t,T\right) }{R_{21}\left( 0,T\right) }=e^{-2\alpha
l}=e^{-2\alpha v_{g}t},  \label{rho21-decay3}
\end{equation}%
where $l$ is the propagation distance and $v_{g}=\Omega _{c}^{2}/2\eta $ is
the group velocity. It is straightforward to obtain $\kappa =4\left(
kv_{s}/\Omega _{c}\right) ^{2}\Gamma $ with eq.($\ref{rho21-decay3}$). In
contrast to the Gaussian decay in eq.($\ref{rho21 decay}$), which depends on
temperature only, the decay of ground-state coherence in SL depends on the
temperature, coupling field and the spontaneous decay rate of level $%
\left\vert 3\right\rangle $. The numerical simulations of $R_{21}\left(
t,T\right) $ and the analytic predictions are plotted for various
temperatures in Fig.6, where good agreements are demonstrated. It is not
unexpected that our numerical simulations closely agree with those obtained
by averaging the solutions of optical-Bloch equations subjected to the
Maxwell-Boltzmann distribution at a particular temperature, since in our
numerical calculations so far, the effect of recoil is negligible, i.e., $%
\left( k_{p}-\tilde{k}_{c}\right) v_{s}\gg \hbar (k_{p}-\tilde{k}%
_{c})^{2}/2m $ [see the definition of $\triangle _{v}$ below eq.($\ref{wave
slowly varying 3}$)]. However, we note that the above criterion is no longer
valid if the mass of atom is made small and the coupling and probe beams are
counter-propagating. The resultant dephasing can significantly reduce the
output level of the probe pulse and our scheme appears applicable to this
kind of problems.

Finally, it should be noted that, simply by adjusting the temperature (or
equivalently, the velocity distribution), we can simulate the decaying
behavior of the ground-state coherence, provided that the explicit
time-dependence of the coupling field is given. This can not be achieved via
solving the optical-Bloch equation by imposing a phenomenological decay rate
$\gamma $ on the metastable ground state $\left\vert 2\right\rangle $ (which
is accurate only at $\Omega _{c}\gg kv_{s}$ in SL and not valid at all in
LS), and thus features the major decoherence between our formalism and the
optical-Bloch equation. For this reason, it is expected that the current
scheme can be applied to investigate the dynamics of stationary light pulse
(SLP) which is basically consisted of four SL processes --- two
co-propagating and two counter-propagating. Owing to the inherent
complexity, specifying the high-order $\gamma $'s in the SLP turns out to be
tricky when solving the usual optical Bloch equations \cite{Pan} . Under the
circumstances, without doubt, our scheme appears to be an easier and more
natural way to approach the problem.

\begin{figure}[htbp]\begin{center}
\includegraphics[width=3.6365in]{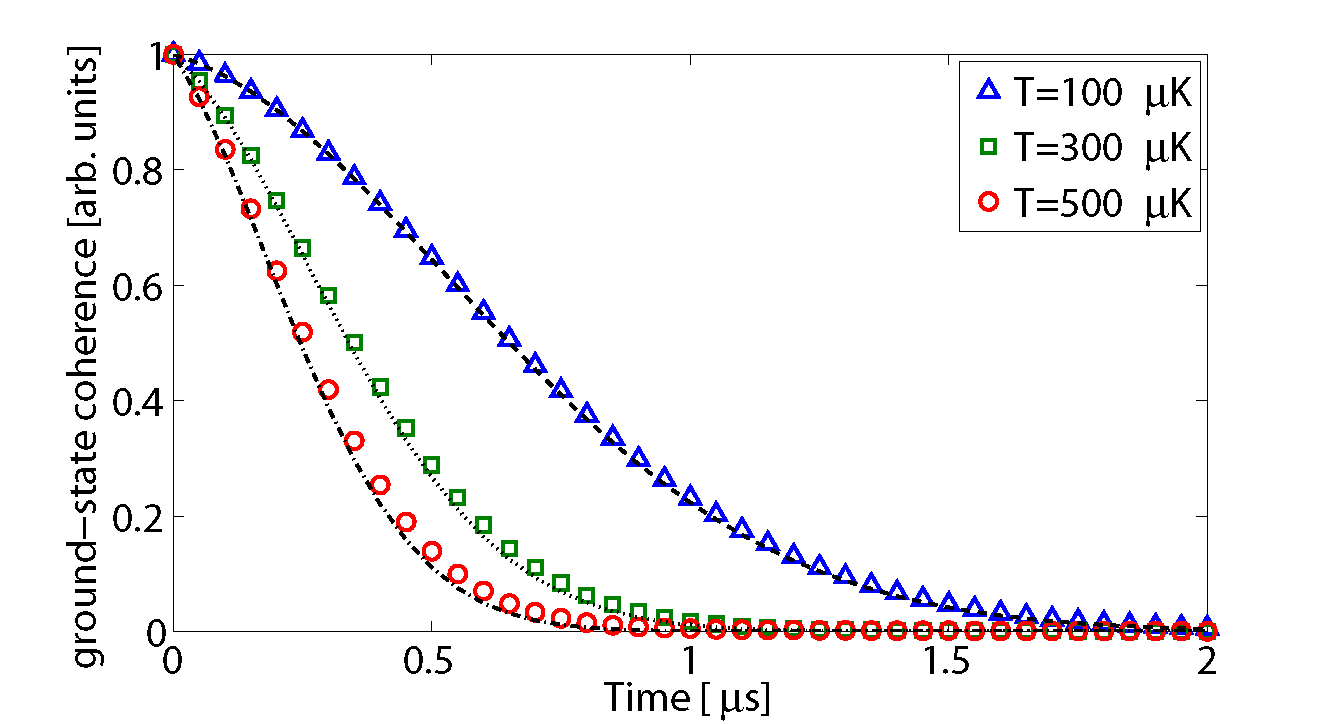}
\caption{(Color online) The triangles,
squares and circles are the numerical results of the normalized $R_{21}(t,T)$
during the storage at $T=100$, $300$ and $500$ $\protect\mu $K,
respectively; the dashed-dotted, dotted and dashed lines are the predictions
by eq.(\protect\ref{rho21 decay}).}
\label{fig5}
\end{center}\end{figure}

\begin{figure}[htbp]\begin{center}
\includegraphics[width=3.6357in]{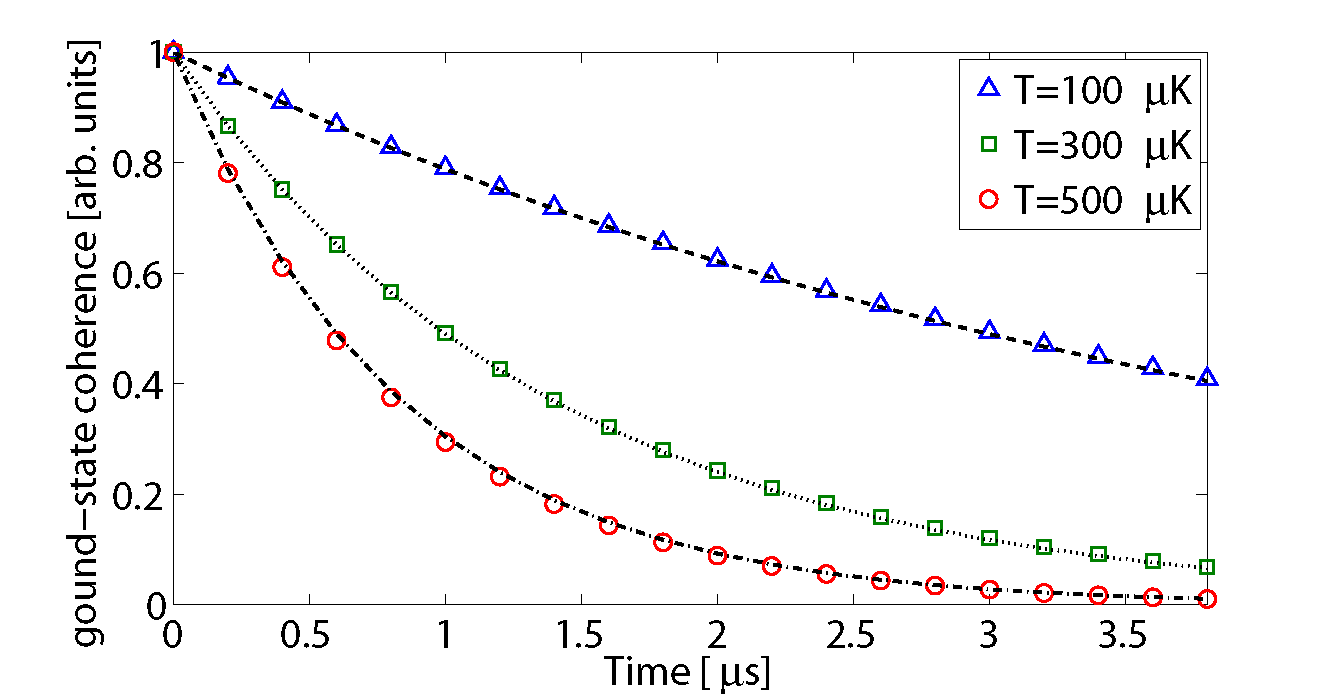}
\caption{(Color online) The normalized $%
R_{21}\left( t,T\right) $ with $\Omega _{c}=$ $0.75\Gamma $, $OD=400$; the
triangles, squares and circles are the numerical results at $T=100$, $300$
and $500\protect\mu $K, respectively; the dashed-dotted, dotted and dashed
lines are the predictions of eq.(\protect\ref{rho21-decay3}).}
\label{fig6}
\end{center}\end{figure}

\section{Concluding remarks}

We have presented a numerical scheme to study the dynamics of SL and LS in
an EIT medium at finite temperatures. Based on the gauge invariance of Schr%
\"{o}dinger equation under Galilean transformation, we derive a set of
coupled equations for a boosted closed 3-level EIT systems. The loss of
ground-state coherence at finite temperatures is then treated as a
consequence of superposition of density matrices representing the EIT
systems moving at different velocities. Unlike other theoretical treatments
in which atoms are assumed immobile, our scheme takes both atom's external
and internal degrees of freedom into full account. The feasibility of this
scheme is shown by comparing the numerical results to the experimental data
for both SL and LS. Last but not least, this scheme also enables us to study
the dynamical properties of a Doppler-broadened EIT medium in the
non-perburbative regime of probe and coupling pulses with comparable
intensities, in which new effects are expected to arise.

\begin{center}
\textbf{ACKNOWLEDGEMENTS}
\end{center}

This work is supported in part by National Science Council, Taiwan under
Grant No. 98-2112-M-018-001-MY2 and 98-2628-M-007-001. TLH and SCG
acknowledge the supports from Taida Institute for Mathematical Science
(TIMS) and the National Center for Theoretical Sciences (NCTS).

\end{document}